\documentclass[twocolumn, prb]{revtex4}

\usepackage{graphicx}
\usepackage{dcolumn}
\usepackage{bm}

\begin{document}

\title{High temperature charge and thermal transport properties of the n-type thermoelectric material PbSe}

\author{John Androulakis}
\affiliation{Department of Chemistry, Northwestern University, Evanston, Illinois, 60208, USA}
\author{Duck-Young Chung}
\affiliation{Materials Science Division, Argonne National Laboratory, Argonne, Illinois, 60439, USA}
\author{Xianli Su} 
\affiliation{Department of Physics, University of Michigan, Ann Arbor, Michigan 48109, USA}
\affiliation{State Key Laboratory of Advanced Technology for Materials Synthesis and Processing, Wuhan University of Technology, Wuhan, 430070, China.}
\author{Li Zhang}
\affiliation{Department of Physics, University of Michigan, Ann Arbor, Michigan 48109, USA}
\affiliation{State Key Laboratory of Advanced Technology for Materials Synthesis and Processing, Wuhan University of Technology, Wuhan, 430070, China.}
\author{Ctirad Uher}
\affiliation{Department of Physics, University of Michigan, Ann Arbor, Michigan 48109, USA}
\author{Thomas C. Hassapis}
\affiliation{Physics Department, Aristotle University of Thessaloniki, 54124, Greece}
\author{Euripides Hatzikraniotis}
\affiliation{Physics Department, Aristotle University of Thessaloniki, 54124, Greece}
\author{Konstantinos M. Paraskevopoulos}
\affiliation{Physics Department, Aristotle University of Thessaloniki, 54124, Greece}
\author{Mercouri G. Kanatzidis}
\email{m-kanatzidis@northwestern.edu}
\affiliation{Department of Chemistry, Northwestern University, Evanston, Illinois, 60208, USA}
\affiliation{Materials Science Division, Argonne National Laboratory, Argonne, Illinois, 60439, USA}


\date{\today}

\begin{abstract}
We present a detailed study of the charge transport, optical reflectivity, and thermal transport properties of n-type PbSe crystals. A strong scattering, mobility-limiting mechanism was revealed to be at play at temperatures above 500 K. The mechanism is indicative of complex electron-phonon interactions that cannot be explained by conventional acoustical phonon scattering alone. We applied the first order non-parabolicity approximation to extract the density of states effective mass as a function of doping both at room temperature and at 700 K. The results are compared to those of a parabolic band model and in the light of doping dependent studies of the infrared optical reflectivity. The thermal conductivity behavior as a function of temperature shows strong deviation from the expected Debye-Peierls high temperature behavior (umklapp dominated) indicating an additional heat carrying channel, which we associate with optical phonon excitations. The correlation of the thermal conductivity observations to the high temperature carrier mobility behavior is discussed. The thermoelectric figure of merit exhibits a promising value of $\sim$ 0.8 at 700K at $\sim 1.5 \times 10^{19}$ cm$^{-3}$.
\end{abstract}

\maketitle

\section{\label{sec:intro} Introduction}
The cubic semiconductor PbSe (space group Fm-3m, a=6.125 \r{A}) has been attracting scientific attention for more that seven decades mainly because of its optical properties. For example, the infra-red behavior of PbSe has been appreciated since the 1940s \cite{Barrow1944} and eventually found applications in photodetectors and thermal imaging.\cite{Ravich1970} More recently, the  advent of nanoscience raised interest in photovoltaic applications based on excitonic effects in PbSe nanocrystals. \cite{Schaller2004, Choi2009, Ma2009} Hence, the compelling majority of published work on the system involves the study of thin films and other nanostructures mainly at room temperature and lower. 

Lately, however, theoretical and experimental reports have pointed out the appealing characteristics of PbSe for higher temperature thermoelectric applications.\cite{Zhang2009, Parker2010, Wang2011, AndroulakisPRB11, AndroulakisJACS11} For example, PbSe melts at a relatively high temperature (1080 $^{o}$C), is composed of earth abundant elements, is easy to scale up, and performs better than PbTe at 900 K, one of the choice thermoelectric materials in the temperature regime 600-900 K.\cite{AndroulakisPRB11,AndroulakisJACS11,Kanatzidis2010} It is noteworthy, that all of the above reports (refs 6-10) have highlighted the lack of detailed experimental studies on the charge and thermal transport as well as basic electronic band structure parameters of PbSe, with different dopants and as a function of doping level especially at high temperatures. 

Hirahara {\it et al.}, for example, studied the mobility of both n and p type PbSe hot pressed samples up to 773 K and doping levels well below $10^{19}$ cm$^{-3}$ taking into account impurity scattering. \cite{Hirahara1954} Later, Schlichting and Gobrecht repeated the mobility study up to 800 K on melt grown crystals of n and p type PbSe extending the doping levels up to $3 \times 10^{19}$ cm$^{-3}$, and concluded that electron phonon interactions are dominant.\cite{Schlichting1973} Scattering was also studied in the framework of defect formation through measurements of Hall effect, electrical conductivity, and thermopower by Gurieva {\it et al.} but the study was not conclusive as to which type of defects (e.g. interstitial, Frenkel etc) dominate.\cite{Gurieva1987} Alekseeva {\it et al.} have studied the high temperature properties of p-type PbSe and samples with isovalent Cd and Mn ion substitution respectively.\cite{Alekseeva} Finally, limited attention has been given to the valence band structure of PbSe and its effect on the high temperature thermoelectric properties.\cite{Vinogradova, Alekseeva1997}

Given the increasing interest in the high temperature properties of PbSe and the aforementioned conflicting conclusions on the charge transport properties of PbSe, it becomes clear that a reliable and consistent experimental body of results has to be established. Hence, we initiated a study of n-type samples doped with Cl exhibiting an electron carrier density in the range $ 7.5 \times 10^{18} \leq n \leq 3.8 \times 10^{19}$ cm$^{-3}$. The choice of Cl as a dopant is justified by its substitutionary action on the Se sublattice leaving undisturbed the conduction band of PbSe that consists primarily of Pb p-orbitals.\cite{Zhang2009, Parker2010} We report electrical conductivity, Hall coefficient, thermoelectric power and thermal conductivity as a function of doping and temperature. Furthermore, the optical reflectivity as a function of doping at room temperature was studied within the framework of a Krammers-Kroning analysis. We employed both a parabolic and a non-parabolic (Kane-type) band model to extract basic parameters such as the effective mass and the Lorenz number at different temperatures and as a function of the doping level. Finally, a thorough investigation of the lattice thermal conductivity is presented where in addition to acoustical and three phonon processes optical phonon contributions are required to better account for the observed high temperature behavior. We find that the maximum thermoelectric figure of merit, ZT, achieved at 700 K is $\sim$0.8 for a carrier density of $\sim 1.5 \times 10^{19}$ cm$^{-3}$.\cite{ZT}

\section{\label{sec:expt} Experimental Details}

The PbSe crystals were grown by the Bridgman technique inside sealed and evacuated quartz ampules with one end tapered. The ampules were loaded with high purity Pb (99.999\%, American Elements), Se (99.999\%, 5N Plus) and PbCl$_{2}$ (99.9999\%, Aldrich). Initially the load was suspended in the hot zone of the furnace at 600 $^{o}$C for 72 h. Then the hot zone was heated to 1150 $^{o}$C, the load was raised at a higher position outside the hot zone and dipped at a speed of $\sim$1.2 mm/h. After growth, the ingots were sliced to 8 mm diameter disks with a waferizing saw. Subsequently, two of the disks were further processes with a polisher to form a bar of typical dimensions $\sim$7$\times$3.5$\times$2.5 mm and to a disk of thickness $\sim$2 mm. Optical examination of the specimens revealed a polycrystalline texture consisting of large single crystals oriented at different directions.

\begin{figure*}[ht]
\centerline{
\includegraphics[width=0.8 \textwidth]{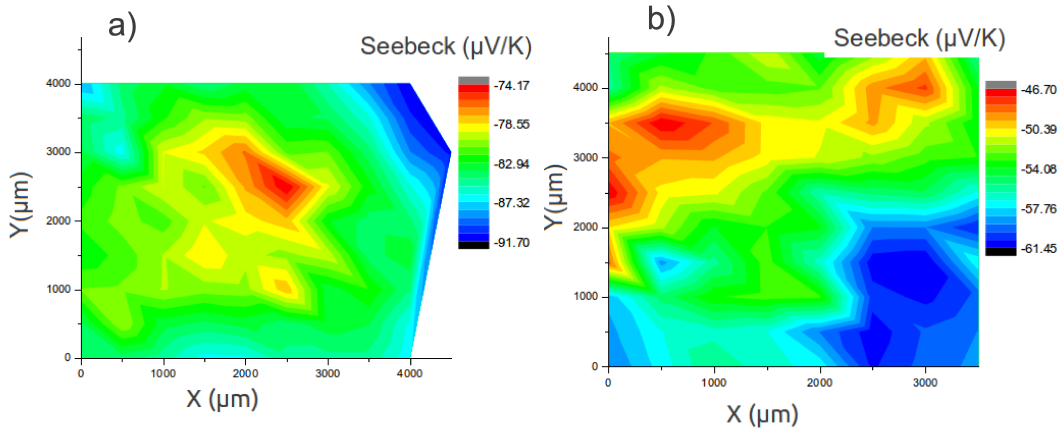}
}
\caption{\label{fig:Figure 1} Scanning Seebeck results on two coins doped with 0.3\% mol PbCl$_{2}$ (a) and 0.5\% mol PbCl$_{2}$ (b). The small inhomogeneous regions randomly observed in (a), which are acceptable for an ingot material, are extended and create steep gradients with increasing PbCl$_{2}$ concentration (b). 
}
\end{figure*}

The bar and disk specimens of each growth were separately examined at room temperature for consistency in doping by combining Hall effect and a spatial scanning Seebeck apparatus. Specimens with large inhomogeneous areas of thermopower, pertaining to large carrier density variations, were not considered further. Figure 1 presents a typical example of a homogeneous and an inhomogeneous pair of specimens. Generally, we have observed that doping levels above 0.4 \% mol, corresponding to an average Hall concentration, $n$, higher than 4$\times 10^{19} cm^{-3}$, produce such high doping inhomogeneities. Therefore, the present study was limited to a set of five samples exhibiting high homogeneity and an average carrier density not higher than 3.8$\times 10^{19} cm^{-3}$, see Table I. The measured $n$ increases monotonically as a function of increasing nominal Cl concentration. A one to one correspondence of $n$ versus Cl concentration was observed to a reasonable degree at $n \leq 1.5 \times 10^{19}$ cm$^{-3}$. At high PbCl$_{2}$ molar concentrations there is a deviation indicating doping action inability of Cl in the PbSe lattice.

The high temperature Hall coefficient was measured in a home-made high temperature apparatus, which provides a working range from 300 K to 873 K. The samples were press-mounted and protected with argon gas to avoid possible oxidization at high temperature.  The Hall resistance was monitored with a Linear Research AC Resistance Bridge (LR-700), and the data were taken in a field of $\pm$ 1 T provided by an Oxford Superconducting air-bore magnet.

The electrical conductivity, $\sigma$, and Seebeck coefficient, S, were measured simultaneously on the bar shaped specimens in a ULVAC-RIKO ZEM-3 system. The specimens were protected in a helium atmosphere ($\sim$ 0.1 atm) while the furnace of the instrument was cycled from room temperature to $\sim$ 700 K and back. No thermal hysteresis was observed with thermal cycling. 

The disk-shaped specimens were used to determine the thermal diffusivity as a function of temperature and doping in a NETZSCH LFA 457 Microflash instrument. Subsequently the thermal conductivity, $\kappa$, was estimated by the relation $\kappa=DC_{p}\rho$, where D is the thermal diffusivity, $C_{p}$ is the heat capacity under constant pressure and $\rho$ is the mass density of the specimens. $C_{p}$ was approximated by the formula $0.171 + (2.65 \times 10^{-5})T$.\cite{AndroulakisPRB11} All charge and thermal transport measurements were performed in the same specimen direction.

\begin{table}[t]
\caption{Specimens of the present study tabulated based on PbCl$_{2}$ content and corresponding carrier, plasmon frequency, and mass density.}
\centering
\begin{tabular}{c c c c c}
\hline
\hline
ID & PbCl$_{2}$ $\% mol$ & n $(10^{19} cm^{-3})$ & $\omega_{P}$ (cm$^{-1}$) & $\rho$ (g/cm$^{3}$)\\
A & 0.05 & 0.75 & 697 & 7.90\\
B & 0.10 & 0.88 & 710 & 7.98\\
C & 0.20 & 1.50 & 729 & 8.02\\
D & 0.30 & 3.20 & 970 & 7.90\\
E & 0.40 & 3.80 & 990 & 7.99\\
F &  -  & $<$0.2 & - & 8.09 \\
\hline
\end{tabular}
\end{table}

Room temperature infrared reflectivity (IR) measurements were performed on finely-polished PbSe samples using a Bruker 113V FTIR spectrometer. The spectra were collected in the 100-3000 cm$^{-1}$ spectral region with a resolution of 2 cm$^{-1}$ at nearly normal incidence. The reflection coefficient was determined by a typical sample-in-sample-out method with a mirror as the reference. The $Im(\epsilon)$ and $Im(-1/\epsilon)$ spectra (where $\epsilon$ is the complex dielectric function) were derived from the Kramers-Kronig transformation.\cite{Fox}

\section{\label{sec:rnd} Results and Discussion}
\subsection{Charge Transport Measurements}

The electrical conductivity, $\sigma$ of samples A-E is depicted in Fig. 2. The values of $\sigma$ are increasing with increasing doping, i.e. moving from specimen A to E, at any given temperature. At room temperature $\sigma$ values as high as 3500 S/cm can be realized for a doping level of 3.8$\times 10^{19}$ cm$^{-3}$, indicative of relatively high mobilities, $\mu$. For all specimens a monotonic decrease in $\sigma$ with increasing temperature is observed. Since $\sigma \approx n \mu e$, the functional dependence of $\sigma$ in temperature may result from the temperature dependence either of $n$ or from factors limiting $\mu$.

\begin{figure}[ht]
\centerline{
\includegraphics[width=0.5 \textwidth]{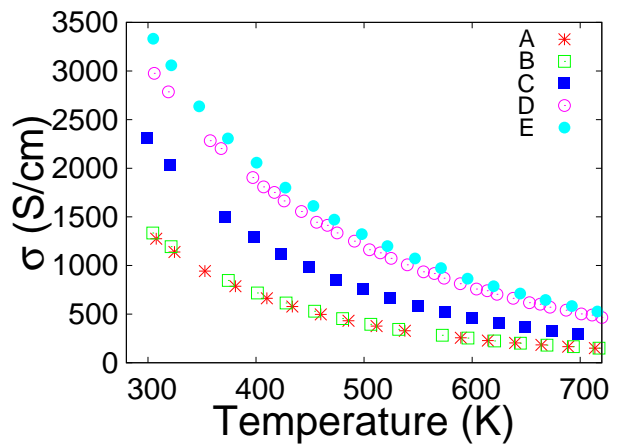}
}
\caption{\label{fig:Figure 2} Electrical conductivity as a function of temperature for samples A-E (see Table I for carrier concentrations). The conductivity increases with increasing doping at any given temperature and no irreversible effects are observed with thermal cycling.  
}
\end{figure}

\begin{figure}[ht]
\centerline{
\includegraphics[width=0.5 \textwidth]{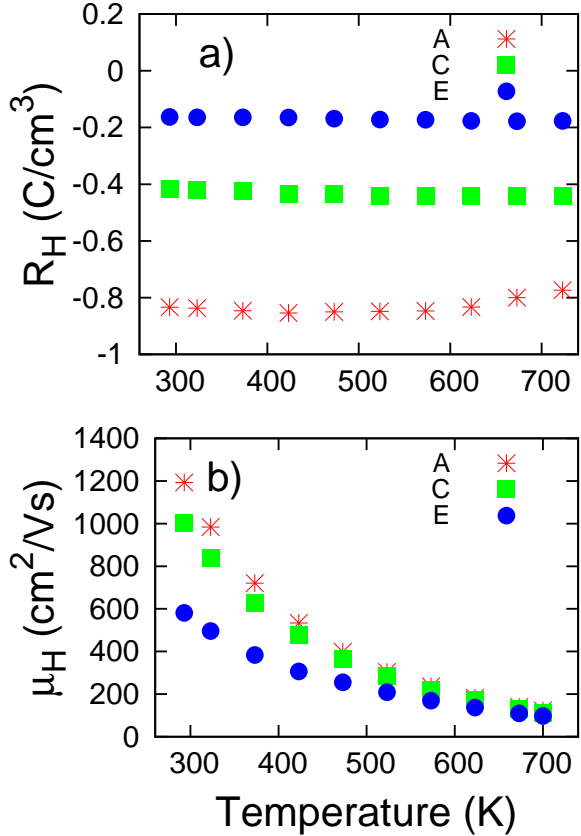}
}
\caption{\label{fig:Figure 3} a) Hall coefficient as a function of temperature for specimens A, C and E. Note that the Hall coefficient is almost temperature independent. b) Hall mobilities for specimens A, C, and E as a function of temperature. Despite the high room temperature values a drastic decrease is observed at high temperatures.
}
\end{figure}

To elucidate the behavior of $\sigma$ we performed temperature dependent Hall effect studies. The Hall coefficient, $R_{H}$, for samples A, C, and E, is plotted in Fig. 3a as a function of temperature. $R_{H}$ is almost temperature independent. Considering $R_{H}=1/ne$ it follows that the specimens retain the same carrier density up to 700 K. A more accurate description of $ R_{H} $ should take into account the non-parabolicity of the bands and the statistical anisotropy due to scattering. In such a case $R_{H}=Ar/ne$,\cite{Abeles1954} where $r$ is the statistical anisotropy factor that varies from 1 to $3 \pi / 8$, and $A$ is the energy surface anisotropy factor that is equal to $A=3K(K+2)/(2K+1)^{2}$.\cite{Schlichting1973} For PbSe, $K=1.75$\cite{Zawadzki1974} and therefore the product $Ar$ varies from 0.97 to 1.14 depending on the value of $ r $. For simplicity we have kept $Ar=1$, since fundamentally the conclusions of our study are not distorted by such an assumption. 

\begin{figure}[ht]
\centerline{
\includegraphics[width=0.5 \textwidth]{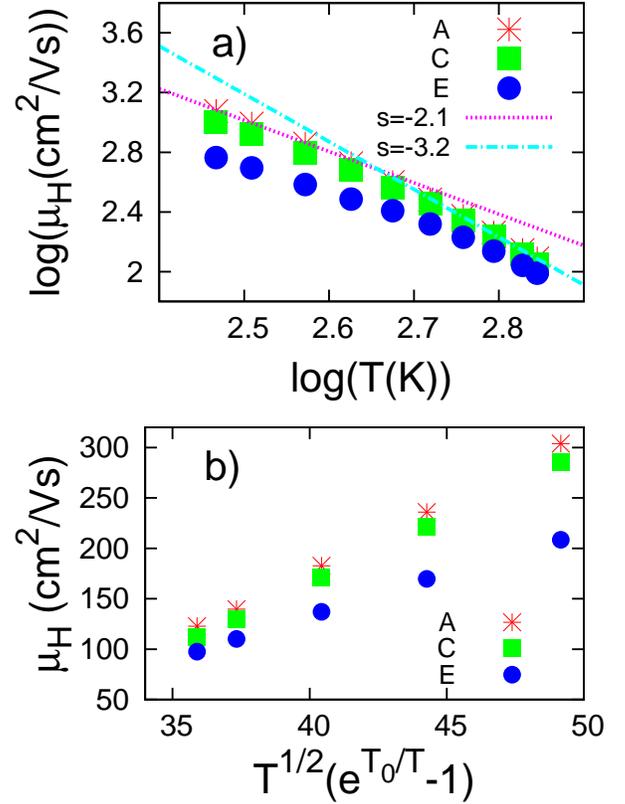}
}
\caption{\label{fig:Figure 4} a) $ log\mu_{H}-logT $ plot showing two regions of linearity one below 480 K and the other above 520 K. The high temperature region is characterized by a much steeper slope reflecting a strong electronic scattering mechanism at play. b) Mobility as a function of temperature scaled according to eq. 1, see text. A linear region is observed that extends from 500 to 700 K. 
}
\end{figure}

Figure 3b, presents the Hall mobility, $ \mu_{H} \approx R_{H}\sigma $, as a function of temperature for specimens A, C, and E. (Specimens B and D exhibit similar behavior but are not included in the discussion for clarity). The room temperature $ \mu_{H} $ values are quite high and drop from $ \sim $1200 cm$^{2}$/Vs for specimen A to $\sim$600 cm$^{2}$/Vs for specimen E. The decrease with doping reflects increasing e-e interactions with band filling. The high values of $ \mu_{H} $ are not retained at high temperatures. A rapid decrease is observed that limits $\mu_{H}$ to $\sim$100 cm$^{2}$/Vs at 700 K. The rapid decrease of $ \mu_{H} $ is clearer on a $ log\mu_{H} - logT $ plot where two linear regions show up, see Fig. 4. In the temperature regime $ 300 \leq T \leq 480 $ K the slopes of the curves are $ \sim $-2. However, for $ T \geq $520K the slopes increase to $ \sim $-3. Table II, summarizes the fitting results. With increasing doping concentration the slopes assume lower values.

\begin{table}[t]
\caption{Characteristic slopes of the log$\mu_{H}$-logT for samples A, C, and E for two different temperature regimes.}
\centering
\begin{tabular}{c c c c c c c c}
\hline\hline
\multicolumn{3}{c}{300$\geq T \geq$480 K} & & &\multicolumn{3}{c}{550$\geq T \geq$750 K} \\
\hline 
 A & C & E & & & A & C & E \\
-2.2 & -2.1 & -1.75 & & & -3.2 & -3.2 & -2.8 \\
\hline
\end{tabular}
\end{table}

Electronic scattering in semiconductors gives rise to distinct temperature dependencies of the carrier mobility. Typical processes considered include: (i) electron-phonon scattering due to thermal vibrations of the lattice, (ii) ionized impurity scattering, (iii) scattering from high frequency vibrations of the lattice (optical phonons), and (iv) scattering by neutral impurities.  On the assumption of parabolic bands, thermal lattice scattering causes $ \mu_{H} $ to scale as $ \mu_{H}^{-1} \sim T^{3/2} $.\cite{Shockley} Non parabolicity of the bands, as in the case of Si,\cite{Pearson} generally increases the value of the temperature exponent to $ \sim $2-2.5.\cite{Mandelis03} In a similar fashion the low temperature exponent in Fig. 4 can be attributed to acoustical phonon scattering of electrons, in agreement with previous results on highly doped PbSe samples.\cite{Schlichting1973} The second mechanism, i.e. ionized impurity scattering, constitutes a positive contribution to mobility\cite{Adachi2005} and is approximated as $ \mu_{H} \sim T^{3/2} $.\cite{Adachi2005} Ionized impurity scattering shows up mostly at low temperatures since the increasing thermal velocity of carriers with increasing temperature effectively screens the Coulomb potential of impurities.\cite{Adachi2005} Finally, mechanism (iv) contributes only weakly to scattering and does not have a significant temperature dependence.\cite{Pearson} Therefore, in order to elucidate the behavior of the mobility at $T>500$ K we have to consider more complex processes such as polar optical phonon scattering. 

The temperature dependence of the electron mobility due to scattering by high frequency optical phonons of the lattice follows the analysis of Fortini et al.,\cite{Fortini1970} that showed a temperature dependence of the form:
\begin{equation}
\mu \approx C T^{1/2} (e^{\eta}-1) G(\eta)
\end{equation}
where, C is a constant of proportionality that involves the static and high frequency dielectric constants and the frequency of the longitudinal optical phonons, $ \eta $ is the reduced chemical potential and $ G(\eta) $ is a function that assumes values from 0.65 to 1.7 depending on the excitation energy of the optical phonons. We point out that Mott and Gurney reached a surprisingly similar result, i.e. a formula containing an activation energy, from a completely different standpoint in the case of ionic polar crystals:\cite{Mott1950}
\begin{equation}
\mu \sim (e^{\Theta/T}-1) 
\end{equation}
where, $\Theta$ is a characteristic temperature between 300 to 800 K. In Fig. 4b we present high temperature mobility data as a function of $T^{1/2}(e^{T_{0}/T}-1)$, where we define $ T_{0} $ as a characteristic activation temperature corresponding to an energy of $ \sim $50 meV. The linear region observed in this case indicates that complicated scattering mechanisms, possibly involving optical phonons, are at play at high temperatures. 

However, the data are not conclusive as to weather an electron-optical phonon scattering is taking place. This is because in PbSe, where the lowest conduction band minimum is at the $ L $ point of the first Brillouin zone (<111> direction),\cite{Ravich1970} and hence is highly degenerate, phonons may scatter electrons transferring them from one valley to another.\cite{Adachi2005} This intervalley scattering mechanism can involve both acoustical and optical phonons and leads to a temperature dependence with similar characteristics as in eq. 1 and 2.\cite{Adachi2005} However, in this case many more parameters are involved such as the intervalley energy separation, the intervalley deformation potential etc.,\cite{Adachi2005} and hence a theoretical study is required to reach a definite conclusion as to the exact nature of the mobility-limiting mechanism at high temperatures. In any case, the data clearly point to deviations from simple electron phonon interactions at temperatures above 500 K and doping levels $ n \geq 7\times10^{18}$ cm$ ^{-3} $. Here we note that Schlichting et al. presented measurements of highly doped $ n \geq 3\times 10^{18}$ cm$ ^{-3} $ PbSe specimens only up to 500 K\cite{Schlichting1973} which may have led to the erroneous conclusion for a simple dominant acoustical phonon scattering mechanism regardless of the doping level.  At lower $n$ and $T>500$K mobility data are plagued by bipolar diffusion which masks other interactions.\cite{Schlichting1973}

The Seebeck coefficient, $ S $ for specimens A-E is presented in Fig. 5 as a function of temperature. For all samples, i.e. for any doping level, $ S $ is almost linearly decreasing from 300 to 700 K assuming higher absolute values with increasing temperature. At 300 K the absolute value of $ S $ is decreasing from sample A to E consistent with increasing carrier density. The same behavior is observed at all temperatures. 

\begin{figure}[ht]
\centerline{
\includegraphics[width=0.5 \textwidth]{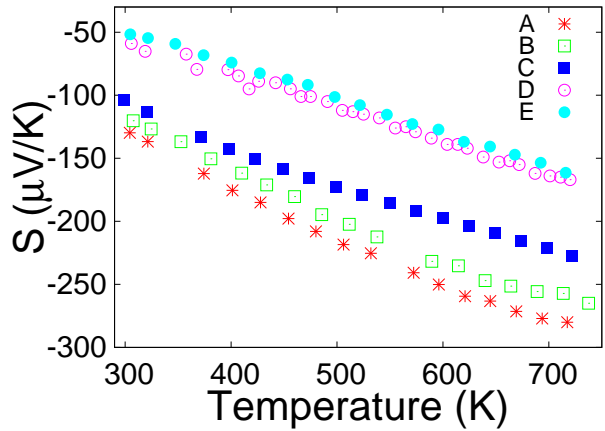}
}
\caption{\label{fig:Figure 5} Seebeck coefficient as a function of temperature for samples A-E. 
}
\end{figure}

Assuming parabolic bands and an energy independent relaxation time then at any temperature  $ S(n) $ is described by a unique effective mass value, $ m^{\ast} $, and for any $n$ the $ S(T) $ depends on the temperature dependence of $ m^{\ast} $.\cite{Johnsen2011, Fistul} The aforementioned assumptions lead to simple formulas that have been explained in detail elsewhere,\cite{Johnsen2011} and can be used to construct Pisarenko plots, i.e. $ S(n) $ diagrams at different temperatures. Such plots are depicted in Fig. 6 for three different temperatures, 300 K, 500 K, and 700 K and the corresponding $m^{\ast}$ values 0.28, 0.35, and 0.41 that allowed fitting of the data (solid lines). The effective mass value is increasing as a result of an increasing band gap, $E_{g}$, with increasing temperature,\cite{Ravich1970} consistent with the results of $ \overrightarrow{k} \cdot \overrightarrow{p}$ theory.\cite{Marder} The temperature dependence of $m^{\ast}$ is plotted in Fig. 7a and compared to $E_{g}(T)$. For PbSe $E_{g}(300 K)\approx0.275$ eV\cite{AndroulakisPRB11} and $ \partial E_{g} / \partial T \approx 4\times10^{-4}$ eV/K.\cite{Smirnov1961} It is evident that in the crude approximation of the parabolic band model $ m^{\ast} $ follows very close the rate of the band gap increase.

As mentioned earlier, the above values are only crude approximations since the electronic band structure of PbSe close to the Fermi level is non parabolic.\cite{Parker2010} Non parabolicity pertains to a non-spherical Fermi surface shape, and hence the dispersion relation depends on the spatial direction. In general, a non parabolic energy band dispersion can be expanded in a power series:\cite{Zawadzki1974}
\begin{equation}
k^{2} =\frac{2m}{\hbar^{2}}(\varepsilon + \sum_{q=2}^{\infty} \lambda_{q}\varepsilon^{q} )
\end{equation}
where the coefficients $\lambda_{q}$ are defined from the following relation:\cite{Zawadzki1974}
\begin{equation}
\lambda_{q} = \frac{\hbar^{2}}{2m} \frac{1}{q!} (\frac{d^{q}k^{2}}{d \varepsilon^{q}})
\end{equation}

\begin{figure}[ht]
\centerline{
\includegraphics[width=0.5 \textwidth]{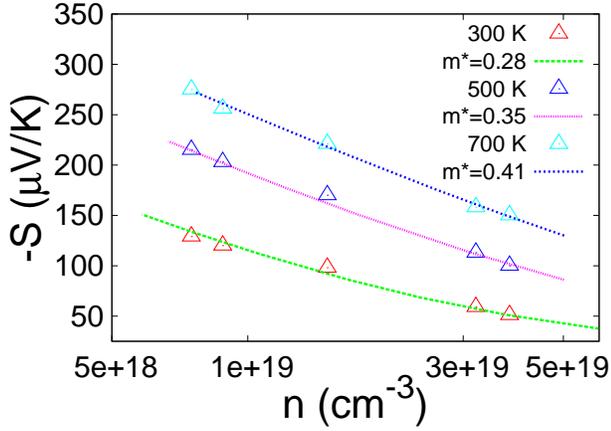}
}
\caption{\label{fig:Figure 6} Pisarenko plots at 300 K, 500 K, and 700 K. The solid lines represent fittings of the data points on the assumption of a parabolic conduction band and an energy independent scattering time. The only adjustable parameter in the calculation was $m^{\ast}$. Note that the higher the temperature the higher the effective mass.  
}
\end{figure}

The first order non-parabolicity approximation ignores all terms with $q \geq 3$ and thus eq. 1 is greatly simplified to: 
\begin{equation}
k^{2} =\frac{2m}{\hbar^{2}} \varepsilon (1 + \lambda \varepsilon)
\end{equation}
In this approximation the coefficient $\lambda$ is usually taken to be equal to the inverse of the band gap, $E_{g}$.\cite{Smirnov1967} Consequently, all galvanomagnetic coefficients can be expressed as functions of the generalized Fermi integrals $^{i}L_{l}^{j}$ defined by the equation:\cite{Zawadzki1974}
\begin{equation}
^{i}L_{l}^{j} (\eta, \beta) =  \int_{0}^{\infty} (- \frac{\partial f_{0}}{\partial z}) z^{i} (z+\beta z^{2})^{j} (1+2\beta z)^{l} dz
\end{equation}
where $f_{0} (\eta, T)$ is the Fermi distribution function, $\eta$ is the reduced chemical potential, $z=\varepsilon/k_{B}T$, and $\beta = \lambda k_{B}T =k_{B}T/E_{g}$. 

\begin{figure}[ht]
\centerline{
\includegraphics[width=0.5 \textwidth]{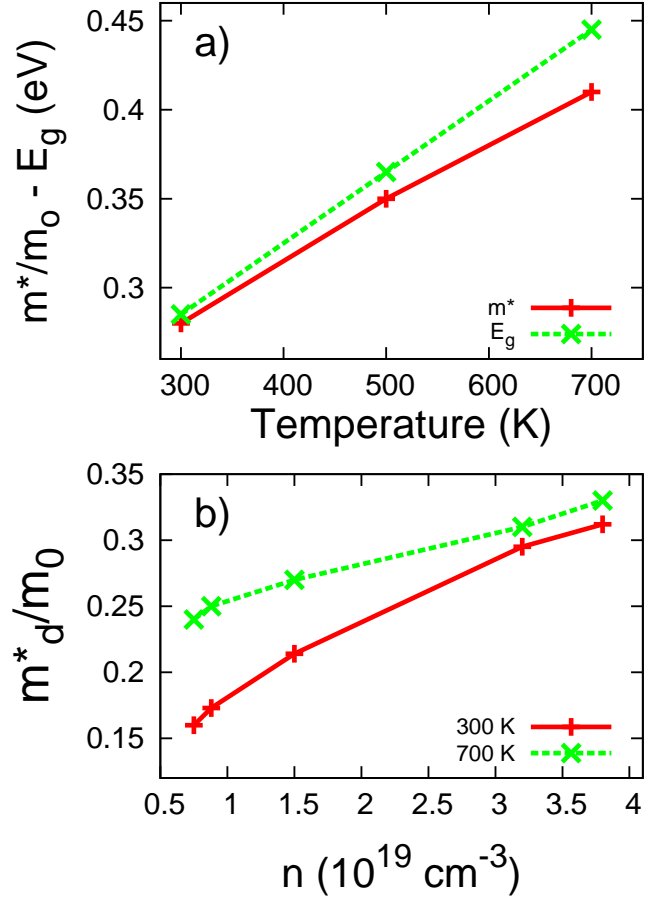}
}
\caption{\label{fig:Figure 7} a) Temperature dependence of the effective mass as calculated from the parabolic band model and comparison with the temperature dependence of the band gap. Note that the results are independent of the carrier density. b) The effective mass calculated from Seebeck coefficient data in the first order non-parabolic approximation.
}
\end{figure}

Here, we are especially concerned with the expression for $S$. Fitting $ S $ yields $\eta$ values that can be utilized in the calculation of other parameters in the same fashion as with assuming a parabolic band.\cite{Johnsen2011} $S$ is defined by:\cite{Zawadzki1974, Smirnov1967}
\begin{equation}
S =  \frac{k_{B}}{e} \frac{{}^{1}L^{1}_{-2} - \eta ({}^{0}L^{1}_{-2})}{{}^{0}L^{1}_{-2}}
\end{equation}

Using the $\eta$ values extracted from fitting the $ S(n) $ data with eq. 7 and in conjunction with the expression for the carrier density (with a unity Hall factor):\cite{Zawadzki1974}
\begin{equation}
n = \frac{1}{3\pi^{2}} (\frac{2m_{d}^{\ast} k_{B} T}{\hbar^{2}})^{3/2} ({}^{0}L^{3/2}_{0})
\end{equation}
the density of states effective mass, $m_{d}^{\ast}$, can be calculated at any given pair of $n$ and $T$ values. The dependence of $m_{d}^{\ast}$ on $ n $ is another substantial difference from the parabolic band model and is supported experimentally in lead chalcogenides.\cite{Dixon1965, Dixon1965b, Aziza1970}

The results of the fitting process using eq. 7 and 8 are depicted graphically in Fig. 7b for 300 K and 700 K. The extracted values are lower compared to those calculated from the parabolic band model, see Fig. 7a, but closer to 0.21, the textbook value for PbSe.\cite{Slack1995} Both at 300 and 700 K $m_{d}^{\ast}$ is monotonically increasing with increasing doping as discussed above.  

\begin{figure}[t]
\centerline{
\includegraphics[width=0.4 \textwidth]{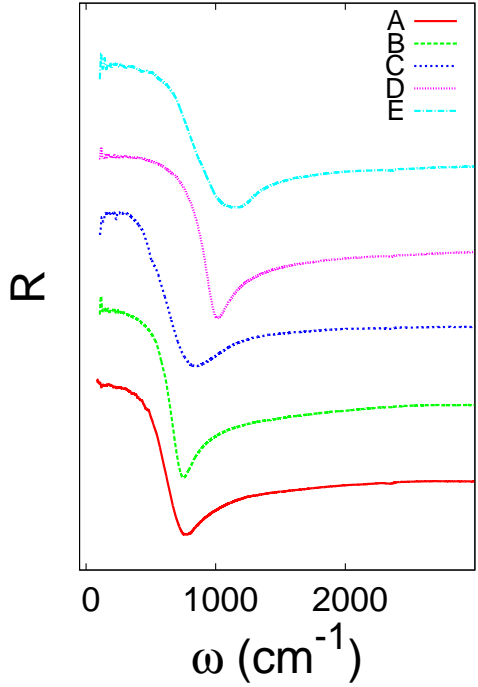}
}
\caption{\label{fig:Figure 8} Room temperature reflectivity spectra as a function increasing frequency. Spectra were shifted vertically by 0.4 for clarity. The reflectivity minimum shifts to higher values from sample A to E as a result of increasing carrier concentration.}
\end{figure}

\begin{figure}[t]
\centerline{
\includegraphics[width=0.5 \textwidth]{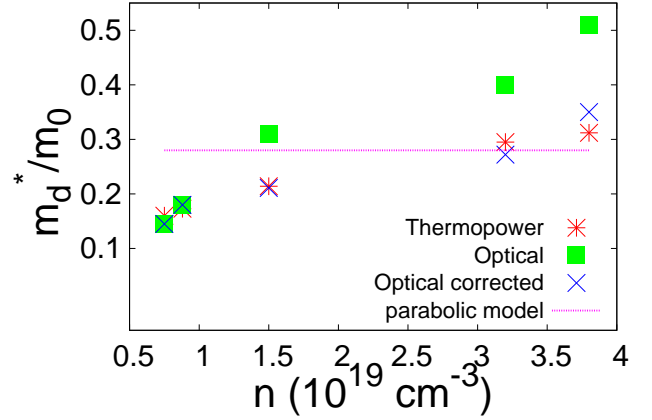}
}
\caption{\label{fig:Figure 9} The density of states effective mass as a function of increasing carrier density extracted from: thermopower measurements analyzed with a single parabolic band model (magenta line), thermopower measurements analyzed in the first order non-parabolicity approximation ($\ast$ marks), optical reflectivity measurements with the assumption that $\epsilon_{3000 cm^{-1}} \approx \epsilon_{\infty}$  (solid green squares), optical reflectivity measurements with a correction of $\epsilon_{\infty} \sim 25$ for all specimens (blue $\times$ marks). 
}
\end{figure}

\subsection{Optical Reflectivity: independent measurement of the effective mass}

The study of the reflectivity as a function of energy in the infrared part of the electromagnetic spectrum  yields useful information about basic materials parameters that are related to the electronic band structure, such as the effective mass.\cite{Fox} This is because the reflectivity, $R$, depends on the contribution of both the bound and the free electrons to the real part of the complex index of refraction. Therefore, we have performed infrared reflectivity measurements in our samples as an independent, yet direct, method of probing the effective mass and thus compare the results with those of the transport models as discussed in the previous section.

The room temperature infra red optical reflectivity as a function of incident radiation wavelength is depicted in Fig. 8 for samples A through E. It is readily seen that the minimum, associated with the plasma frequency, $\omega_{P}$ and hence the carrier concentration, is steadily increasing from specimen A to E reflecting the increasing doping. Accordingly, the Kramers-Kronig transformation yielded blue-shifted absorption peaks in $Im(1/\epsilon)$, in the region $>$700 cm$^{-1}$, that were used to accurately determine the plasma frequency, $\omega_{P}$. The transverse optical - longitudinal optical splitting in the reflectivity spectra of PbSe occurs in the 34-114 cm$^{-1}$ region.\cite{Burhkard1974} The latter makes plasmon-phonon effects negligible in the present study. Therefore, the reflectivity minima in Fig. 8 are mainly determined by the contribution of free carriers.

The plasma frequency is related to basic materials parameters, such as the electric susceptibility (or optical) effective mass, $m_{op}^{\ast}$, through the relation:
\begin{equation}
\omega_{P}^{2}=\frac{ne}{\epsilon_{\infty} \epsilon_{0} m_{op}^{\ast}}
\end{equation}
where $n$ is the carrier concentration, $\epsilon_{\infty}$ is the high frequency dielectric constant (a measure of the bound electron contributions to the dielectric function),$\epsilon_{0}$ the vacuum permeability, and $e$ is the electron charge. The expression for the dependence of R on frequency, $\omega$, at nearly normal incidence:
\begin{equation}
R(\omega)=(\frac{ \sqrt{\epsilon(\omega)}-1 }{\sqrt{\epsilon(\omega)}+1})^{2}
\end{equation}
where $\epsilon(\omega)$ is the complex dielectric function. Using the experimental R values at 3000 cm$^{-1}$ and eq. 10 we calculated the value of $\epsilon_{\infty}$ ($\epsilon_{\infty} \approx \epsilon_{3000cm^{-1}}$). We have observed that $\epsilon_{\infty}$ is decreasing from 25 to 17 with increasing n. However, there is no physical reason for a changing contribution of the bound electrons with n in PbSe, and thus, we have taken $\epsilon_{\infty} \sim$ 25 for all samples. 

The values of $\omega_{P}$ (see Table I) were derived from the peak value of $Im(-1/\epsilon)$ obtained by Kramers-Kronig transformation. Using eq. 9, with the Hall-effect extracted $n$ we have evaluated $m_{op}^{\ast}$. The calculated values of $m_{op}^{\ast}$ can be transformed to $m_{d}^{\ast}$ by using the relation:\cite{Zawadzki1974,Aziza1970}
\begin{equation}
m_{d}^{\ast}=N_{m}^{2/3}\frac{1+2K}{3K^{3/2}}m^{\ast}_{op}
\end{equation}
where $N_{m}$ is the number of equivallent conduction band ellipsoids in the first Brillouin zone and $K$ is the ellipsoid anisotropy factor which for PbSe takes the value $K=1.75$.\cite{Zawadzki1974} Our results are plotted in Fig. 9 as the green squares and compared to the extracted $m_{d}^{\ast}$ values from thermopower data analysis ($\ast$ marks). We observe that the optical measurements support a strongly increasing $m_{d}^{\ast}$ with increasing $n$ in agreement with the non-parabolic nature of the conduction band of PbSe. The agreement between the optically extracted and charge transport extracted $m_{d}^{\ast}$ values is excellent for $\epsilon_{\infty} \sim 25$ (blue $\times$ marks). Obviously, the divergence of the data at high $n$ are associated with an underestimation of the $\epsilon_{\infty}$ values with heavy doping.

\begin{figure}[t]
\centerline{
\includegraphics[width=0.5 \textwidth]{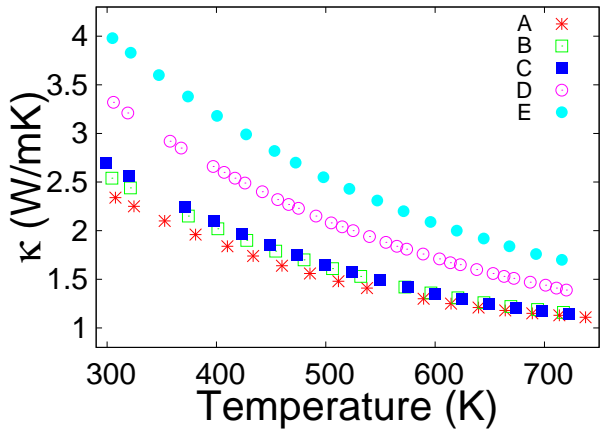}
}
\caption{\label{fig:Figure 10} Total thermal conductivity as a function of temperature.
}
\end{figure}

\subsection{Thermal Transport}

The total thermal conductivity ($ \kappa = \kappa_{e} + \kappa_{lat} $, where $ \kappa_{lat} $ is the lattice part and $ \kappa_{e} $ the free carrier contribution) as a function of temperature is presented in Fig. 10. The room temperature values start from $ \sim $2.4 W/mK for sample A and increase with increasing doping to $ \sim $4 W/mK for sample E. Rising temperature increases the electron-phonon and phonon-phonon interactions which cause $ \kappa $ to decrease. Interestingly, $ \kappa $ remains above 1 W/mK at all temperatures, despite the strong mobility-reducing mechanism that is in effect above 500 K and should also limit the heat carrying efficiency of carriers. Therefore, we conclude that the latter may result from an increased lattice contribution at high temperatures, i.e. another mechanism contributing to heat conduction. 

$ \kappa_{lat} $ is estimated indirectly by use of the Wiedemann-Franz relation, $\kappa_{e}=L \sigma T$, where $L$ is the Lorenz number and subtraction from the total. The temperature dependence of $L$ is critical in the proper calculation of $\kappa_{e}$. A good approximation that yields reasonable results is the assumption of a parabolic band, electron acoustic phonon interaction only and a constant relaxation time.\cite{Johnsen2011, AndroulakisPRB11} Figure 9a presents the results of such a calculation for samples A, C, and E. The calculated Lorenz number values are below the metallic limit ($L_{0} \approx 2.45 \times 10^{-8}$ W$\Omega$/$K^{2}$) and decrease with decreasing doping and increasing temperature. 

In the first-order non parabolic approximation the Lorenz number is expressed:\cite{Zawadzki1974}
\begin{equation}
L =  (\frac{k_{B}}{e})^{2} (\frac{{}^{2}L^{1}_{-2} {}^{0}L^{1}_{-2} - ({}^{1}L^{1}_{-2})^{2}}{({}^{0}L^{1}_{-2})^{2}})
\end{equation}
where ${}^{i}L_{l}^{j}$ are the integrals defined by eq. 6 and are functions of $\eta$. The temperature dependence of $\eta$ is extracted by fitting $S(T)$ data, see Fig. 5, with eq. 7. The results are plotted as a function of temperature in Fig. 11b. Comparing with the results of the parabolic model, it is evident that the first order non-parabolicity leads to higher Lorenz number values for the same doping level. At high doping (sample E) the room temperature value is slightly larger than that of the metallic limit, $ L_{0} \approx 2.45\times 10^{-8} W \Omega / K^{2}$. This may be due to an inadequacy of the non-parabolicity approximation ($ \lambda \sim E_{g}^{-1} $) at high doping. Both models, however, exhibit essentially the same functional dependence with respect to temperature and therefore the temperature dependence of $ \kappa_{lat} $ remains the same.

\begin{figure}[t]
\centerline{
\includegraphics[width=0.5 \textwidth]{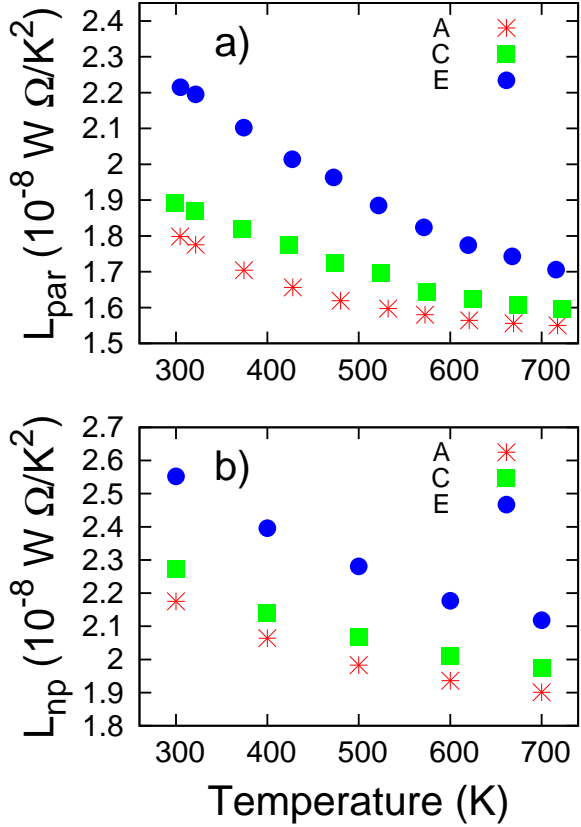}
}
\caption{\label{fig:Figure 11} a) Lorenz number for samples A, C, and E calculated as a function of temperature on the assumption of a parabolic conduction band, electron-acoustical phonon interactions only and constant relaxation time. b) Same as in a) however the first order non-parabolicity approximation is used, see text.
}
\end{figure}

Due to the overestimation of the Lorenz number of the non-parabolic model at high doping we have used the predictions of the parabolic model to extract lattice thermal conductivity as a function of temperature. The results are depicted graphically in Fig. 12a. For clarity we use the results for samples A and E only. Curves for samples B-D have similar values and temperature dependence, leading us to conclude that $ \kappa_{lat} $ is independent of doping consistent with the low concentration of PbCl$_{2}$ used in this study. 

Generally, the temperature behavior of  $ \kappa_{lat} $ when umklapp processes are dominant is $ \kappa_{lat} \sim T^{-1}$. This is the case of PbTe.\cite{Alekseeva1983} In the case of PbSe, however, we find that $ \kappa_{lat} ~T^{1-\delta} $ with $ \delta \approx 0.2-0.23 $. In Fig. 12b $ \kappa_{lat} $ is presented as a function of $ 1000/T^{0.8} $. The solid line is a linear fit of the data confirming the scaling behavior. In order to exclude such a behavior stemming from fitting artifacts (Lorenz number calculation) or from doping, despite the dilute Cl concentration, we grew separately a pure, undoped PbSe single crystal, here referred to as sample F in Table I. (The same crystal was also used in a previous study \cite{AndroulakisPRB11}). The carrier concentration in the undoped crystal was determined through Hall measurements to be $ < 2\times10^{18} cm^{-3}$. In such a case the contribution of free carriers is minimized. In confirmation of the scaling behavior mentioned above the $\kappa_{lat}$ of specimen F exhibits a similar temperature dependence with $\delta \approx 0.23$ and a room temperature value of $\sim$1.9 W/mK.

\begin{figure}[t]
\centerline{
\includegraphics[width=0.5 \textwidth]{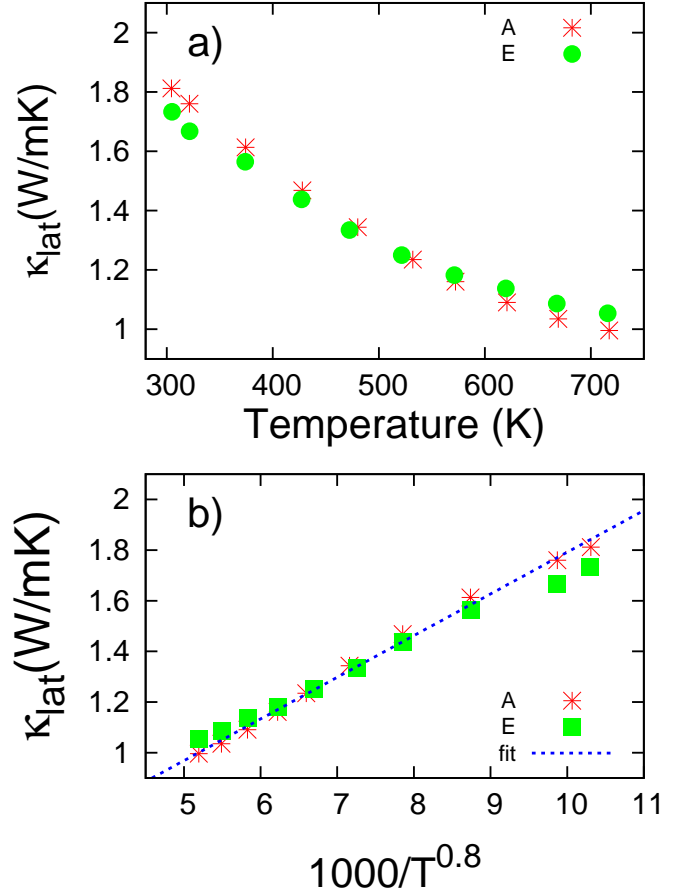}
}
\caption{\label{fig:Figure 12} a) Lattice thermal conductivity extracted using the $L_{par}$, see Fig. 11, temperature dependence. Notice that $ \kappa_{lat} $ is doping independent. b) Scaling of the lattice thermal conductivity as $ \sim T^{0.8} $. The dashed line is a linear fit (sample A), used here as a guide to the eye. 
}
\end{figure}

Usually a positive $ \delta $ is attributed to optical phonon excitations that provide an extra heat carrying path.\cite{Steigmeier1963, Steigmeier1966, Alekseeva1983, Hess2004} This raises the thermal conductivity at high temperatures compared to the simple $\delta=0$ behavior. Steigmeier and Kudman, used the well known result of three-phonon processes at high temperatures:
\begin{equation}
\kappa_{lat} \sim (\frac{k_{B}}{h})^{3} \frac{\alpha^{4}\rho \theta_{D}^{3}}{\gamma^2T}
\end{equation}
in combination with temperature dependent $ \kappa_{lat} $ accurate data of several III-V compounds received with the flash diffusivity - heat capacity method.\cite{Steigmeier1963} They concluded that eq. 13 (where, $ k_{B}, h, \alpha, \rho, \theta_{D}, \gamma $ are the Boltzmann constant, Planck's constant, the lattice parameter, the density, the Debye temperature, and the Gr\"{u}neisen parameter respectively) is valid only when $\gamma$ is temperature dependent, indicating the presence of optical phonons contributing to heat conduction. A similar analysis on our PbSe specimens leads to the same conclusion on $\gamma$. Interestingly, a more simplified model developed by Dugdale et al.\cite{Dugdale1955} with a similar physical basis as the three phonon model, was previously found adequate to describe the temperature dependence of $ \kappa_{lat} $ on the assumption of a temperature dependent $\gamma$.\cite{AndroulakisPRB11} 

To separate the contribution of acoustical phonons from the $\kappa_{lat}$ we have calculated the phonon thermal conductivity by assuming a phonon scattering relaxation time $\tau(x)$\cite{Callaway1960}
\begin{equation}
\tau(x)^{1}=\tau_{D}^{-1}+\tau_{P}^{-1} = A\omega^{4} + C T \omega^{2}
\end{equation}
where we have considered only contributions from point defects (D index) and umklapp processes (P index). In the above $\omega$ is the phonon frequency. Hence the acoustical phonon contribution to $\kappa_{lat}$ takes the form:
\begin{equation}
\kappa_{l,ac} = \frac{k_{B}}{2\pi^{2} \upsilon_{S}} (\frac{k_{B}T}{\hbar})^{3} \int_{0}^{\theta_{D}/T}  \frac{x^{2}}{Dx^{2}+E'} \frac{e^{x}}{(e^{x}-1)^{2}} dx
\end{equation}
where $ \upsilon_{s} $ the speed of sound, $x=\hbar \omega/k_{B}T$ the dimensionless variable of the phononic energy, $D$ is a temperature independent constant and $E'=CT(k_{B}T/\hbar)^{2}$. Since, optical phonons presumably appear at temperatures above $\theta_{D}$ we have used literature\cite{Ravich1970, Shalyt1968} $\kappa_{lat}$ values at $T<\theta_{D}=170$K\cite{Ravich1970}, i.e. at temperatures where only acoustical phonons dominate, to extract the constant $C$. Subsequently $E'$, and $\kappa_{l,ac}$ where calculated. The results are depicted in Fig. 13a along with $\kappa_{lat}$ of sample F.\cite{AndroulakisPRB11} Evidently, there is a considerable deviation which increases with rising temperature. 

\begin{figure}[t]
\centerline{
\includegraphics[width=0.5 \textwidth]{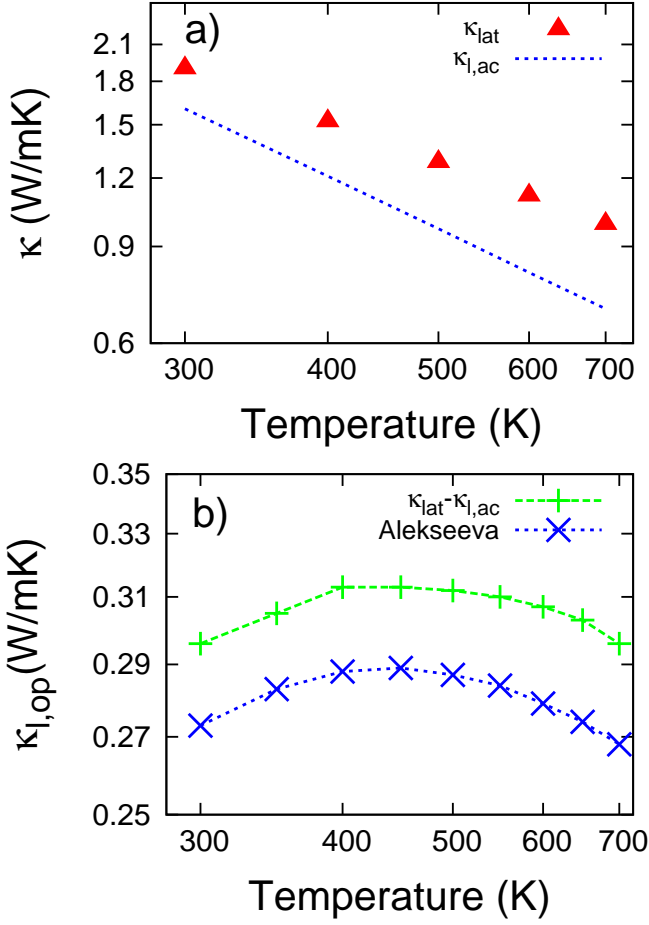}
}
\caption{\label{fig:Figure 13} a) Temperature dependence of the lattice thermal conductivity of PbSe (specimen F, Table I, red triangles) compared to the expected behavior on the assumption of umklapp processes only (dashed blue line). b) The contribution of optical phonons extracted from the difference of the curves in a) (green crosses) and also by fitting the measured PbSe lattice data of specimen F to eq. 16 (blue $\times$ symbols), see text for details. Remarkably, the dependence of $\kappa_{l,op}$ on temperature is the same regardless of the model used.
}
\end{figure}

Interestingly, Alekseeva et al.\cite{Alekseeva1983} have previously observed similar thermal conductivity scalings on the lighter lead chalcogenides, i.e. PbSe and PbS. The authors formulated the relationship:\cite{Alekseeva1983}
\begin{equation}
\kappa_{lat} = \kappa_{l,ac} + \kappa_{l,op} = \frac{a'}{T} + \beta(e^\frac{\hbar \omega_{0}}{k_{B}T} - e^\frac{\hbar \omega_{1}}{k_{B}T})
\end{equation}
where $\kappa_{l,op}$ is the contribution of optical phonons to the lattice thermal conductivity, $a'$ and $\beta$ are fitting constants, and $\omega_{0}$ and $\omega_{1}$ are the minimum and maximum optical phonon frequencies. Equation 16 was extracted based on experimental phonon spectra of lead chalcogenides.\cite{Alekseeva1983}

The temperature dependence of the contribution of optical phonons can be calculated either by the subtraction $\kappa_{lat}-\kappa_{l,ac}$, see eq. 15, or directly fitting $\kappa_{lat}$ with Alekseeva's model (eq. 16). Using data received on sample F we have employed both models. Alekseeva's model resulted in $\hbar \omega_{0} \sim 12.4$ meV, $\hbar \omega_{1} \sim 89.4$ meV, and $\beta \sim 0.47$ W/mK. (Note that $T_{0}$ in Fig. 4b amounts to $\sim$ 50 meV, i.e. the average of $\hbar \omega_{0}$, $\hbar \omega_{1}$ minimum and maximum optical phonon energies as suggested by eq. 16). The contribution of optical phonons to the total lattice thermal conductivity as a function of temperature is graphically depicted in Fig. 13b. A comparison of the results gives a rough 7 \% disagreement between the values predicted by the two models at all temperatures, with Alekseeva's model lying lower. Nevertheless, a striking similarity in the functional form of $\kappa_{l,op}$ with respect to temperature is observed, where a broad peak value appears at $\sim$450 K. It is interesting that this temperature correlates with the transition temperature of the electronic Hall mobilities (see Fig. 4a) progressing from a predominant acoustical phonon scattering region (300-450 K) to a region of stronger, possibly optical phonon, scattering (T$>$520 K). We would like to point out that the peak is not related to a maximum contribution of optical phonons. Since the contribution from acoustical phonons is rapidly decreasing ($\sim T^{-1}$) the percentage contribution of the optical phonons to the total thermal conductivity is constantly  increasing with increasing temperature giving rise to the $T^{1 - \delta}$ dependence.

The physical origin of the optical phonon contributions identified here is currently unknown but it may be associated with the increasing displacement of Pb atoms from the octahedron center in the rock salt structure discovered recently in PbQ (Q=S, Se, Te).\cite{Bozin2010} Interestingly, even in the heaviest lead chalcogenide, i.e. PbTe, anharmonic contributions are present in both the charge and the thermal transport properties. Feit et al.\cite{Feit1983} identified significant polar optical phonon contributions in n-type PbTe and more recently Delaire et al.\cite{Delaire2011} identified a strong coupling of the transverse optic mode with the acoustical longitudinal mode which is believed to overall keep the thermal conductivity low.\cite{Delaire2011} However, in contrast to PbS and PbSe the lattice thermal conductivity of PbTe exhibits a 1/T behavior at high temperatures, i.e. $\delta = 0$.\cite{Alekseeva1983} The above emphasize that despite their striking structural similarity and simplicity, lead chalcogenides present challenging electronic structure complexity that varies from Te to Se and likely S. In any case despite the increasing experimental evidence for significant participation of interactions involving high frequency out of phase, i.e. anharmonic, phonons in the heat conduction process in a wide and diverse variety of compounds\cite{AndroulakisJACS11,Steigmeier1966,Alekseeva1983,Hess2004,Delaire2011,Morelli2008} leading to either an increase or a decrease of $\kappa_{lat}$ there is still no adequate theoretical treatment. 

\begin{figure}[t]
\centerline{
\includegraphics[width=0.5 \textwidth]{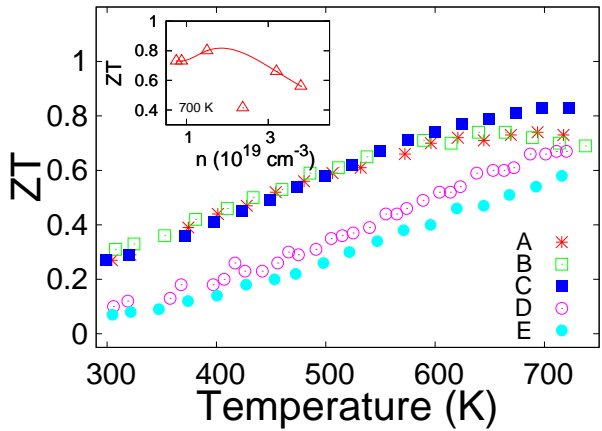}
}
\caption{\label{fig:Figure 14} The thermoelectric figure of merit as a function of temperature for all samples. The maximum ZT at 700 K is observed $\sim 1.5 \times 10^{19}$ cm$^{-3}$, inset.
}
\end{figure}

The ZT, \cite{ZT} is depicted as a function of temperature in Fig. 14. The maximum value (0.8 at 700 K) is assumed for sample C (n$\sim 1.5 \times 10^{19}$ cm$^{-3}$). Given the high temperature mobility reducing mechanism, the ZT  values reached here emphasize that PbSe is a promising thermoelectric material. We note that chemical substitutions on the Se sublattice with inexpensive and highly abundant S have led to ZTs as high as 1.3 at 900 K.\cite{AndroulakisJACS11}

\section{Concluding Remarks}

We have performed a detailed study of the charge transport and thermal transport properties of n-type, Cl-doped PbSe. A strong mobility-limiting mechanism, most probably related to polar optical phonon scattering of free carriers, was shown to be in operation at high temperatures. Thermal conductivity analysis identified an extra heat carrying path in PbSe in the form of polar optical phonon excitations related to the above mobility reducing mechanism. Applying a single parabolic band model with a constant relaxation time results in oversimplifications and therefore should be applied with caution. The first order non-parabolic model, for carrier concentrations $< 1 \times 10^{19}$ cm$^{-3}$ and approximating the non-linear coefficient with the inverse of a linearly increasing band gap, was shown to describe satisfactorily basic electronic structure parameters of PbSe such as the density of states effective mass. The latter was independently extracted by optical reflectivity measurements and found to be in good agreement with the charge transport results. At high temperatures and higher carrier densities a better non-parabolic approximation of the conduction band is necessary. Finally, our results indicate a great potential of PbSe for applications as a thermoelectric material at high temperatures. The involvement of optical phonons in conducting heat at high temperatures imply that in PbSe the lattice thermal conductivity at 700-900 K is higher in relative terms than in PbTe where optical phonons are less important. Therefore, strategies for reducing thermal conductivity to raise ZT should take these findings into account and be aimed at creating optical phonon scattering mechanisms.

\subsection*{Acknowledgements}
The authors would like to acknowledge sponsorship and scanning Seebeck measurements by ZT Plus Thermoelectric Materials (www.ztplus.com). This work was also supported as part of the Revolutionary Materials for Solid State Energy Conversion, an Energy Frontier Research Center funded by the U.S. Department of Energy, Office of Science, 
Office of Basic Energy Sciences under Award Number DE-SC0001054. The work at Argonne National Laboratory is supported by Department of Energy, Office of Basic Energy Sciences (Grant No. DE-AC02-06CH11357).


\begin{thebibliography}{53}
\bibitem{ZT} ZT is defined as $ZT=\frac{S^{2} \sigma T}{\kappa}$, where $S$ is the Seebeck coefficient, $\sigma$ the electrical conductivity and $\kappa$ the total thermal conductivity.
\end{thebibliography}
\end{document}